# Insights into the structural, electronic and optical properties of $X_2MgZ_4$(X= Sc, Y; Z= S, Se) spinel compounds: Materials for the future optoelectronic applications

G. Murtaza

[1]*Materials Modeling Lab, Department of Physics, Islamia College University, Peshawar*

 **Abstract**

Direct bandgap bulk materials are very important for the optical applications. It is therefore important to predict new materials with the desired properties. In the present work, density functional theory is applied to study different physical properties of $X_2MgZ_4$(X= Sc, Y; Z= S, Se) spinel compounds. Generalized gradient approximation is used to analyze the structural and elastic parameters while modified Becke Johnson exchange potential is applied to calculate electric band profiles and optical properties. All the studied compounds are stable in the cubic structure. Also the energy bandgap is of direct nature. Therefore these compounds can find useful applications in the optoelectrics devices. Optical properties of the compounds are studied in terms of dielectric function, refractive index, extinction coefficient, optical conductivity and reflectivity. The transport parameters like electrical conductivity, Seebeck coefficient, and thermal conductivity are also evaluated.

\*Corresponding author: E-mail address: kmwong123a@gmail.com

1. **Introduction**

Semiconductors are the back bone of optoelectronic industry. The efficient optoelectronic devices can be grown with the selection of suitable energy bandgap in a particular energy range of the electromagnetic spectrum. The suitable energy bandgap is the direct valence to conduction energy gap for the efficient optoelectronic. It is therefore natural to predict new materials with the desired properties to meet the demands of the rapidly growing technology.

The $A_2BX_4$ spinel materials, with $A$ and $B$ metallic elements and $X$ from chalcogen (O, S, Se, Te) family, have attracted high attention [1-2], since it put forward a resourceful range of significant physical properties i.e. phase transformations [3], transparent character for longer energy range [4], birefringence [5], high photosensitivity [4], nonlinear optical susceptibility [5], colossal magnetoresistance [6], half metallicity [7], metal insulator transitions [8], humidity sensing [9], high catalytic activity [10], charge storage [11], thermoelectricity [12] etc., which make them potential candidates for manufacturing of optoelectronic devices and materials for numerous applications in geophysics, magnetism, catalysis and environment [13-20].

Ab initio calculations offer one of the most powerful tools for carrying out theoretical studies of an important number of physical and chemical properties of the condensed matter. Therefore, in the present work $X_2MgZ_4$(X= Sc, Y; Z= S, Se) spinel compounds are investigated for the first time using density functional theory (DFT). The work is important due to its novelty and it brings forth new materials which have suitable properties for the use in advance technology. Further this work will motivate the experimentalist to synthesise these compounds and characterize.

2. **Methodology**

The calculations corresponding to the $X_2MgZ_4$(X= Sc, Y; Z= S, Se) compounds are performed with the known precise scheme for solving the Kohn-Sham equations with the full potential linearized augmented plane wave (FP-LAPW) [21] method including the local orbitals (lo) to treat semicore states [22] as implemented in Wien2k [23]. The radii of the muffin tin spheres are 2.2, 2.24, 2.45, 2.14, and 2.41 bohr for Sc, Y, Mg, S, and Se respectively. For the computational analysis and optimization of the structural properties of the materials, the exchange-correlation functional is treated with the generalized gradient approximation (GGA) parameterized by Perdew Burke Ernzerhoff [24]. On the other hand, the new semi-local potential known as the "Tran and Blaha modified Becke-Johnson" TB-mBJ [25] potential, has been employed for determining the electronic and optical properties.

**3. Results and discussion**:

*3.1. Structural and dynamical parameters:*

The magnesium based $X_2MgZ_4$(X= Sc, Y; Z= S, Se) compounds have cubic spinel structure shown in Fig. 1. Volume optimization is a usual route to extract the important structural properties of materials through first principles calculations. In this route, the unit cell volume varied in a particular range and corresponding change in the unit cell energy of the crystal is noted. The energy data corresponding to the volume (E-V) is fitted by the Birch Murnghan equation of state and shown in the Fig. 2. It is visible from the curves that have a parabolic pattern verifies the optimization of the unit cell volume. The bottom point of the curve represents the ground state energy and volume because it has the ground state density. Different structural parameters obtained through these calculations are presented in table 1. It is seen from the table that the unit cell volume and lattice constant increases by changing the cation from Sc to Y and the anion from S to Se. Bulk modulus of the compounds is also decreased by changing the cation from Sc to Y and the anion from S to Se. The similar trend of variation of lattice constant and bulk modulus was also determined earlier in Ref. [26] for other compounds too. No structural data regarding these compounds have been reported so far. Therefore this data can be useful for further theoretical and experimental studies.

Structural stability of a system can be verified by the calculations of the elastic moduli. The cubic systems require only three elastic constants ($C_{11}$, $C_{12}$, and $C_{44}$) to characterize them fully. The cubic crystals are elastically stable, when it meets the following Born stability conditions [27];

$$C_{11} - C_{12} > 0;\ C_{11} + 2C_{12} > 0;\ C_{11} > 0;\ C_{44} > 0$$

The calculated elastic constants of the compounds are also presented in table 1. It is seen that all compounds fulfil the stability conditions. It is seen from the table that $C_{11}$ is much larger than $C_{44}$, so these compounds show high resistance to the unidirectional compression compared to shear deformation.

**3.2. Electronic properties:**

Electronic band structure of the $X_2MgZ_4$(X= Sc, Y; Z= S, Se) spinels is depicted in Fig. 3. It is seen from the figure that top of the valence band and bottom of the conduction band located at the Γ symmetry point. Therefore all the studied compounds are direct bandgap semiconductors. The calculated fundamental bandgaps of the compounds are 2.39 eV, 1.81 eV, 2.78 eV and 2.17 eV for $Sc_2MgS_4$, $Sc_2MgSe_4$, $Y_2MgS_4$, and $Y_2MgSe_4$ respectively. It is inferred that the energy bandgap increases by changing the cation from Sc to Y, while it decrease by replacement of the anion from S to Se. To date no experimental work is reported on these compounds. Therefore this work can be used as a reference data for the experimentalist to explore these direct bandgap materials ranging in the visible region of electromagnetic spectrum and therefore have the ability to absorb solar radiations.

Density of states (DOS) of the compounds $X_2MgZ_4$(X= Sc, Y; Z= S, Se) are shown in Fig. 4. The band below the Fermi level (0 eV) is valence band (VB). In the valence band S/Se-$p$ states contribute majorly throughout the VB alongwith the notable contribution of Sc/Y -$d$ state in the middle of the band. Therefore pd hybridization is present in the valance band. In the conduction band (CB) above the Fermi level, Sc/Y -$d$ states occupied the lower part of the CB, while S/Se -$p$ and $d$ states contribute majorly in the upper part.

The effective mass of electrons ($m_e^*$) is calculated from the band structure of the spinel compounds $X_2MgZ_4$(X= Sc, Y; Z= S, Se), these values are estimated from the curvature of the conduction band minimum. The diagonal elements of the effective mass tensor, $m_e$, for the electrons in the conduction band are calculated using the following well-known relation:

$$\frac{1}{m_e^*} = \frac{1}{\hbar^2} \frac{\partial^2 E(k)}{\partial k^2} \ldots\ldots\ldots\ldots (1)$$

The effective mass of electron is assessed by fitting the electronic band structure to a parabolic function Eq. (1). The calculated electron effect mass ($m_e^*$) of the compounds is decreased in going from S to Se and Sc to Y. The calculated hole effect mass ($m_h^*$) of the compounds is decreased in going from S to Se and Y to Sc. Comparatively the electron effect mass ($m_e^*$) of the compounds is less than the hole effect mass ($m_h^*$).

Electron density distribution reveals the chemical bond nature among the materials. Here the electrons density of the compounds is shown along <110> plane in Fig. 5(a-d). It is observed that for $Sc_2MgZ_4$ (Z= S, Se), Mg-S and Mg-Se bonds are ionic in nature because there is electrons density shared among the Mg-S/Se as visible from the thermometer scale. On the other hand, Sc-S and Sc-Se bonds are covalently bonded due to the charge sharing among them. However the Sc-S/Se bonds are covalent but they are weak in magnitude as clear from the scale and DOS plots in which Sc-d state contribute less compared to S/Se –p state. Similar trend is seen for the $Y_2MgZ_4$ (Z= S, Se) compounds. However the covalent bonding in the $Y_2MgZ_4$ (Z= S, Se) is less than $Sc_2MgZ_4$ (Z= S, Se).

In the transport phenomena, electrons play major rule. Both the electrical energy and heat energy can be transported by electrons. It is interesting to see the response of materials as a function of rise in temperature. Fig. 6 demonstrates the variation of the electronic conductivity per constant relaxation time ($\sigma/\tau$) as a function of temperature. The electrical conductivity increases almost linearly with the increase of temperature because it directly related with the carrier concentration. The carrier concentration increases with the rise in temperature, due to the count of thermally produced carriers. Comparatively $Y_2MgS_4$ has higher electrical conductivity while $Sc_2MgSe_4$ have lower conductivity among the compounds, but the variation trend with temperature is similar among all the compounds. The calculated maximum values of $\sigma/\tau$ at 800 K of the compounds are $8.46\times10^{18}$ (WmS)$^{-1}$, $7.25\times10^{18}$ WmS)$^{-1}$, $9.08\times10^{18}$ WmS)$^{-1}$ and $8.23\times10^{18}$ WmS)$^{-1}$ for $Sc_2MgS_4$, $Sc_2MgSe_4$, $Y_2MgS_4$ and $Y_2MgSe_4$, respectively. The electrical conductivity of the $X_2MgZ_4$ compounds is less compared to the typical thermoelectric materials [28]. This is due to the high energy bandgap of the compounds and alternative confirmation of the wide bandgap nature of these compounds

The thermal conductivity per constant relaxation time ($\kappa/\tau$) in the temperature range 250 – 800 K of $X_2MgZ_4$(X= Sc, Y; Z= S, Se) is shown in Fig. 7. Thermal conductivity ($\kappa$) increases with the rise in temperature, however the increment at higher rate higher in thermal conductivity compared to electrical. The $Y_2MgS_4$ compound has highest thermal conductivity while $Sc_2MgSe_4$ have lowest conductivity among the compounds. The maximum thermal conductivities ($\kappa$) are observed at 800 K of values $4.69\times10^{14}$ (W/m.K.s)$^{-1}$, $4.26\times10^{14}$ (W/m.K.s)$^{-1}$, $4.73\times10^{14}$ (W/m.K.s)$^{-1}$ and $4.35\times10^{14}$ (W/m.K.s)$^{-1}$ for $Sc_2MgS_4$, $Sc_2MgSe_4$, $Y_2MgS_4$ and $Y_2MgSe_4$, respectively. The thermal conductivity of electrons in the $X_2MgZ_4$ compounds is less compared to the typical thermoelectric materials [28]. It is due to the high energy bandgap of these materials.

## 3.3. Optical properties

Dielectric function is an important parameter to characterize the optical properties of materials [29]. It comprises on the real ($\varepsilon_1(\omega)$) and imaginary ($\varepsilon_2(\omega)$) parts which corresponds to the polarization and energy loss in the optical medium, respectively. The calculated $\varepsilon_1(\omega)$ and $\varepsilon_2(\omega)$ are shown in Fig. 8. From the figure it is visible that the compounds zero frequency limits $\varepsilon_1(0)$ varies inversely to the energy band gap i.e. decrease in band gap results in the increase of $\varepsilon_1(0)$. Also $\varepsilon_1(0)$ increases by replacing the anion from S to Se and decreases by replacing the cation from Sc to Y. In the infrared region, the $\varepsilon_1(\omega)$ spectra stays almost constant, in the visible region there is sharp increase in the spectra. It oscillates twice before sharp fall in the ultraviolet region. Even it goes below zero in some particular energy ranges.

Imaginary part $\varepsilon_2(\omega)$ of the dielectric function is also shown in Fig. 8. These compounds are wide bandgap semiconductors, therefore the interband transitions play major role while the intraband transitions are ignored in these calculations. Further, the indirect transitions are also ignored due to their minute absorption ability compared to direct transitions. The spectra starts due to interband transition of the electrons from the upper part of valence band to the lower part of the condition band (S/Se-p states to Sc/Y-d states) with the threshold points at 2.38 eV, 1.82 eV, 2.85 eV, and 2.18 eV for $Sc_2MgS_4$, $Sc_2MgSe_4$, $Y_2MgS_4$, and $Y_2MgSe_4$ respectively. Beyond the thresholds points the spectra sharply grow and prominent peaks appeared due to the high joint density of states. First prominent peak for $Sc_2MgSe_4$ is in the visible region while all the other peaks of the entire compounds lie in the ultraviolet region of the electromagnetic spectrum. It is further noted that the peak value of $\varepsilon_2(\omega)$ increases by replacing the anion from S to Se and decreases by replacing the cation from Sc to Y in all the compounds. Also it is observed that there are two main peaks in the $Sc_2MgS_4$ while three main peaks are seen for the other compounds. It is due to the splitting of the lower part conduction band as seen from the Fig. 3.

The refractive index ($n(\omega)$) and reflectivity ($R(\omega)$) of the compounds $X_2MgZ_4$(X= Sc, Y; Z= S, Se) is shown in Fig. 9. From the figure it is visible that the compounds zero frequency limits, $n(0)$, varies inversely to the energy band gap. Also $n(0)$ increases by replacing the anion from S to Se and decreases by replacing the cation from Sc to Y in these compounds. Beyond the zero frequency limits the spectra starts increasing slowly, however the sharp increase observed in the visible region, it oscillates twice before sharp fall in the ultraviolet region. Even it goes below unity in some particular energy ranges. The maximum refractive index is noted for $Sc_2MgSe_4$.

The frequency dependent reflectivity is also shown in Fig. 9. The zero frequency reflectivity, R(0), of the compounds is more than 13%. It increased by replacing the anion from S to Se and decreased by replacing the cation from Sc to Y in these compounds. Like the refractive index and dielectric function, zero frequency reflectivity also decreases by increase in the energy bandgap. The reflectivity of the compounds increases beyond the zero frequency limits and becomes high in the visible region and prominent in the ultraviolet region. The maximum reflectivity of the compounds is less than 40% in the optical frequency region.

The optical conductivity ($\sigma(\omega)$) and absorption coefficient ($\alpha(\omega)$) of the compounds $X_2MgZ_4$(X = Sc, Y; Z = S, Se) is shown in Fig. 10. From the Fig. 10 it is visible that all the compounds starts conduction in the visible region with critical points at energy 2.385 eV, 1.82 eV, 2.75 eV , and 2.2 eV for $Sc_2MgS_4$, $Sc_2MgSe_4$, $Y_2MgS_4$, and $Y_2MgSe_4$ respectively. Beyond the critical points conductivity sharply increase due to the direct bandgap nature of the compounds. The compounds reveal maximum optical conductivity in the ultraviolet region of the electromagnetic spectrum. Its magnitude decreases by replacing the anion from S to Se and increases by replacing the cation from Sc to Y in these compounds. The maximum conductivity is noted for $Y_2MgS_4$.

The frequency dependent absorption coefficient '$\alpha(\omega)$' is also shown in Fig. 10. The absorption coefficient of the compounds started from 2.35 eV, 1.80 eV, 2.90 eV and 2.18 eV for $Sc_2MgS_4$, $Sc_2MgSe_4$, $Y_2MgS_4$, and $Y_2MgSe_4$ respectively. It is observed that the compounds show high absorption per unit length in the ultraviolet region. The maximum value of the $\alpha(\omega)$ for the compounds is $185\times10^4 cm^{-1}$, $164\times10^4 cm^{-1}$, $154\times10^4 cm^{-1}$, $147\times10^4 cm^{-1}$ in the energy range 0 -15 eV.

## 4. Conclusions

First principles calculations were performed for the first time using FPLAPW+*lo* method to determine the physical properties of $X_2MgZ_4$(X= Sc, Y; Z= S, Se) spinel compounds. All the compounds are stable in the spinel structure. Structural parameters were also determined. The energy bandgap of the studied compounds is of direct nature. Therefore these compounds are optically active and can be potential candidates for the optoelectrics applications. Electrical and thermal conductivities of the compounds increase with the rise in temperature. Optical properties of the compounds demonstrate high optical response of the compounds in the ultraviolet region.

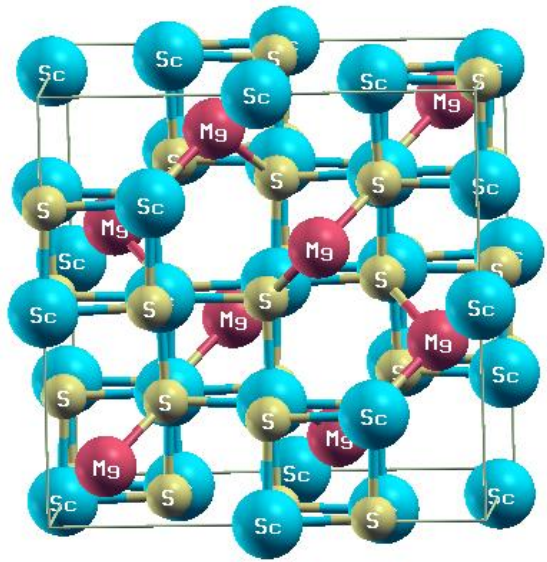

**Fig. 1**. Crystal structure of $Sc_2MgS_4$ as a prototype for $X_2MgZ_4$(X= Sc, Y; Z= S, Se).

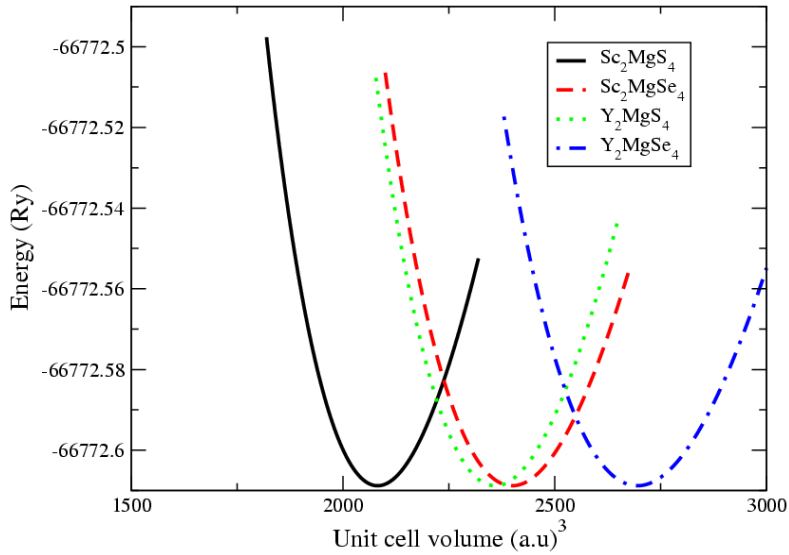

Fig. 2. It is showing the volume optimization of the compounds. All the Energy versus volume (E-V) curves is plotted together by taking the energy difference of the ground state of the highest value. The exact ground states of the compounds are shown in the table 1.

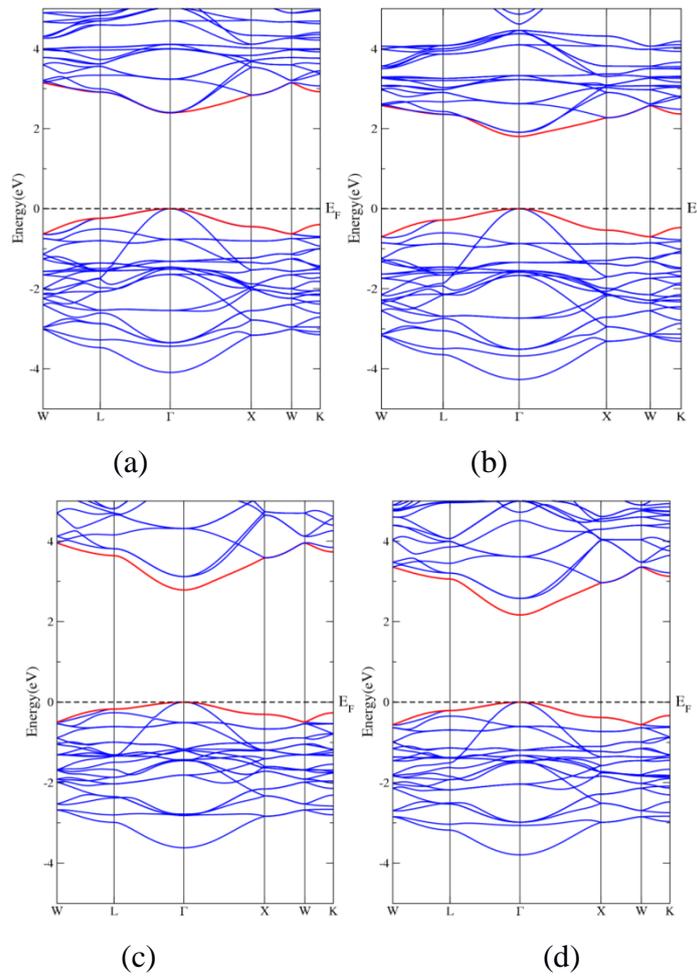

(a)                 (b)

(c)                 (d)

**Fig. 3**. The electronic band structure of the (a) $Sc_2MgS_4$, (b) $Sc_2MgSe_4$, (c) $Y_2MgS_4$, and (d) $Y_2MgSe_4$ compounds.

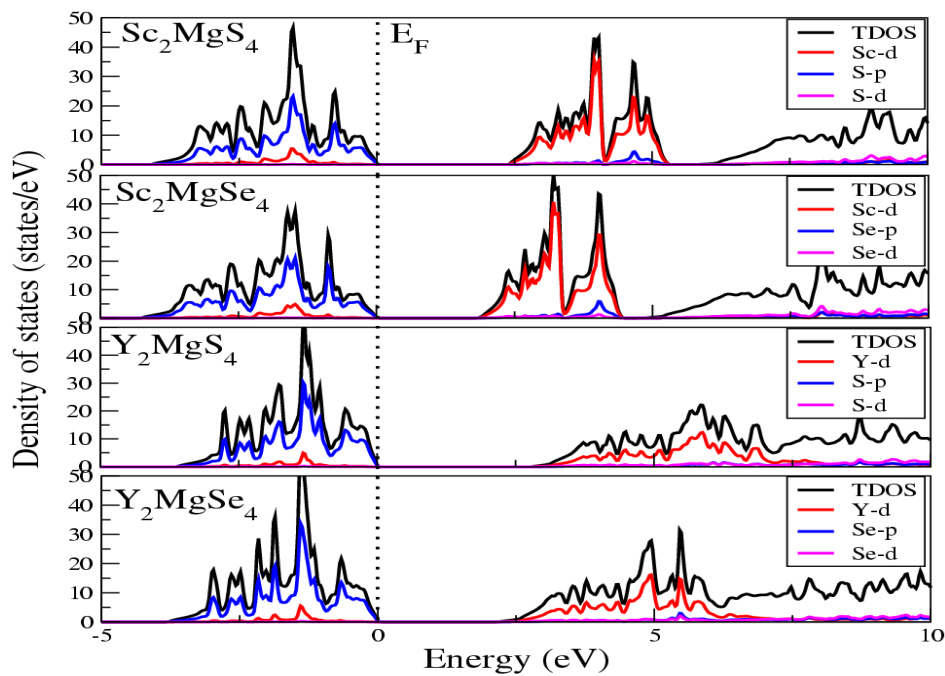

**Fig. 4**. The total and partial density of states of the $X_2MgZ_4$(X= Sc, Y; Z= S, Se) compounds

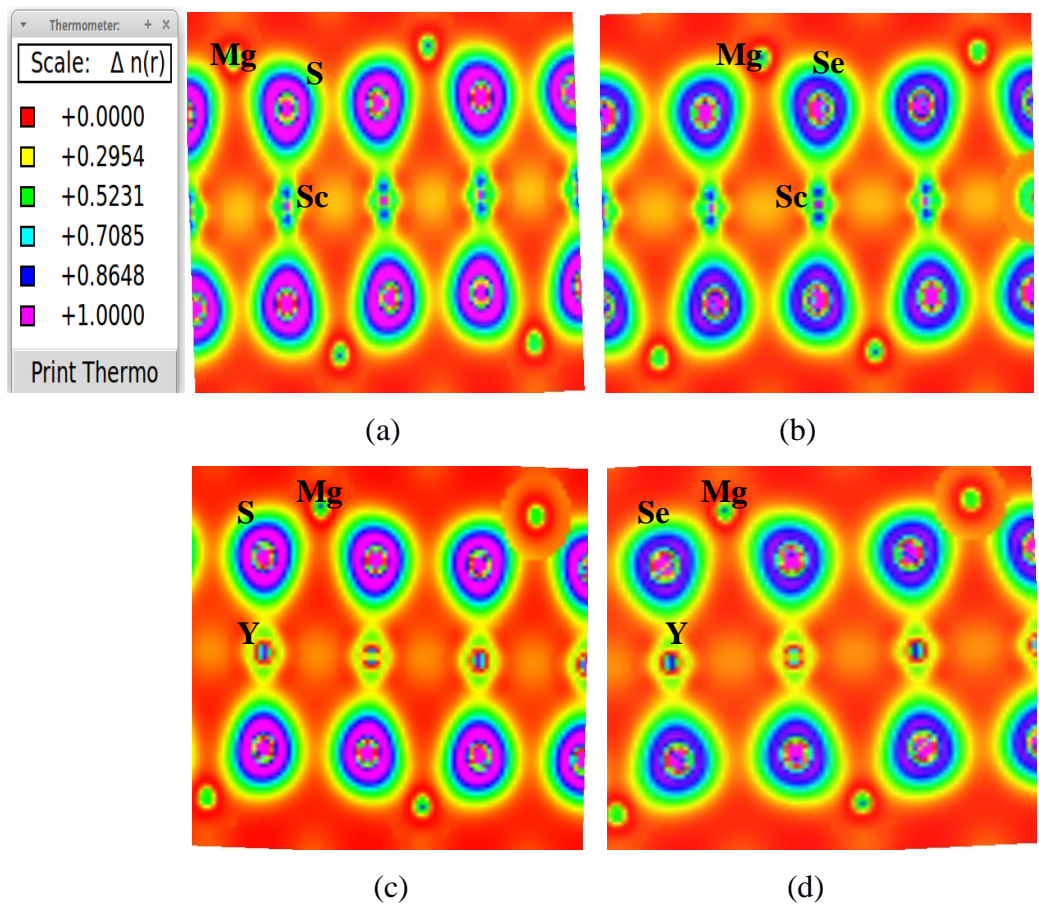

(a)          (b)

(c)          (d)

**Fig. 5**. The electronic charge density of the (a) $Sc_2MgS_4$, (b) $Sc_2MgSe_4$, (c) $Y_2MgS_4$, and (d) $Y_2MgSe_4$ compounds.

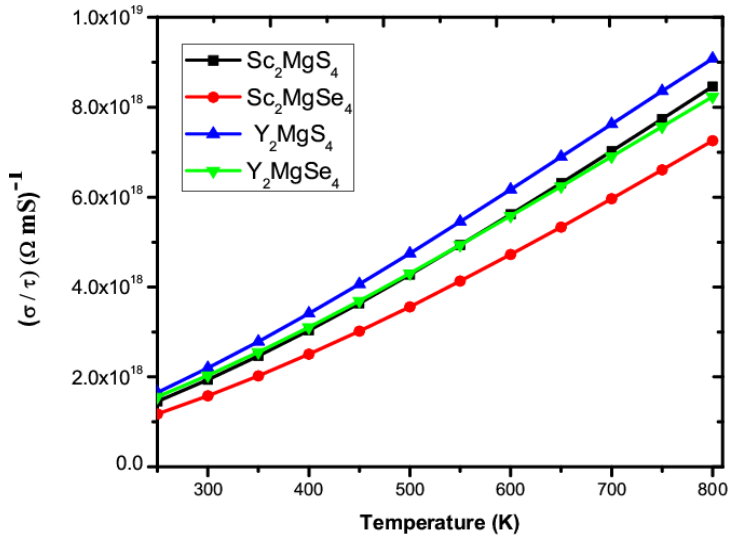

**Fig. 6:** Plot of electrical conductivity versus temperature of $X_2MgZ_4$ compounds.

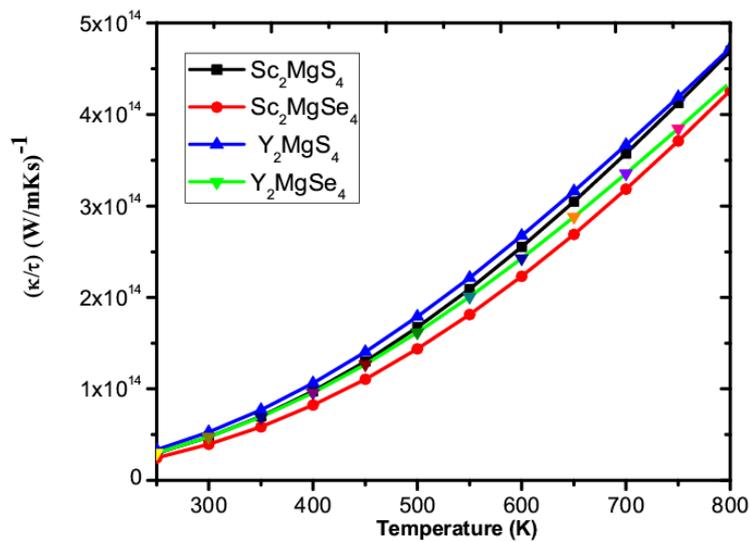

**Fig. 7:** Thermal conductivity versus temperature of $X_2MgZ_4$ compounds.

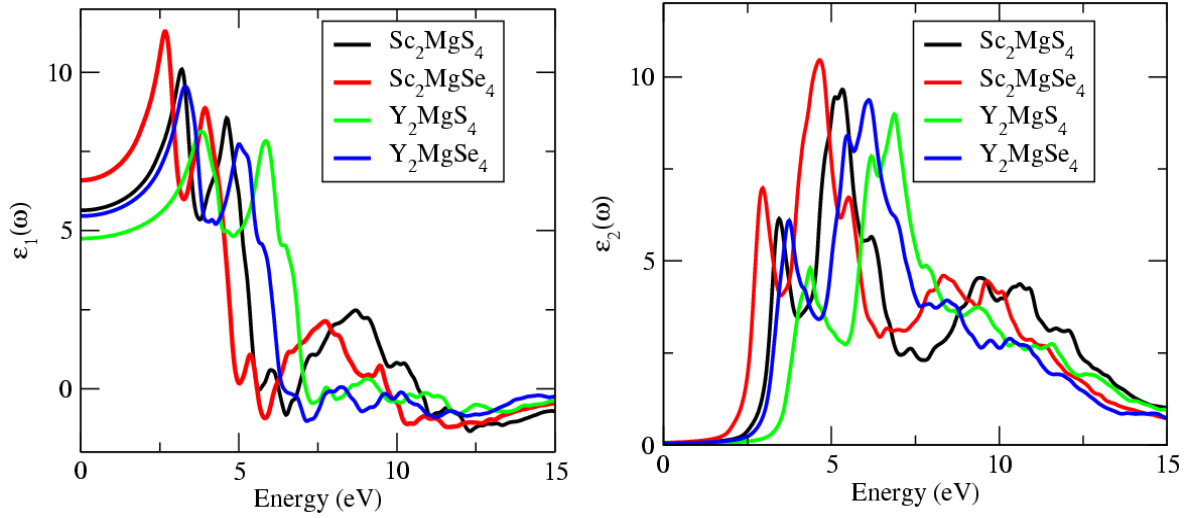

**Fig. 8.** The real ($\varepsilon_1(\omega)$) and imaginary ($\varepsilon_2(\omega)$) parts of the dielectric function as a function of photon frequency.

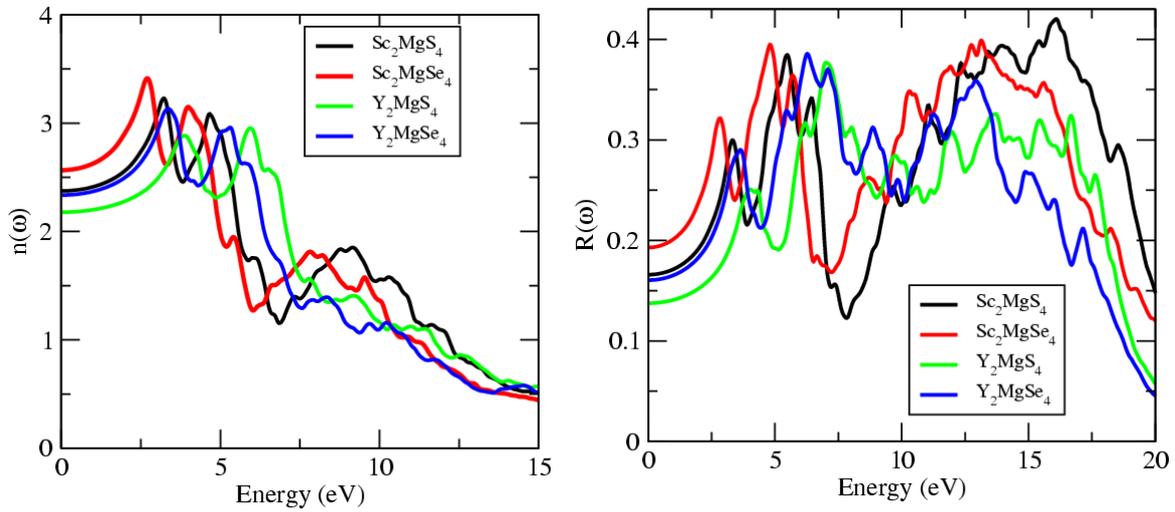

**Fig. 9.** The refractive index and reflectivity of the compounds $X_2MgZ_4$(X= Sc, Y; Z= S, Se) is

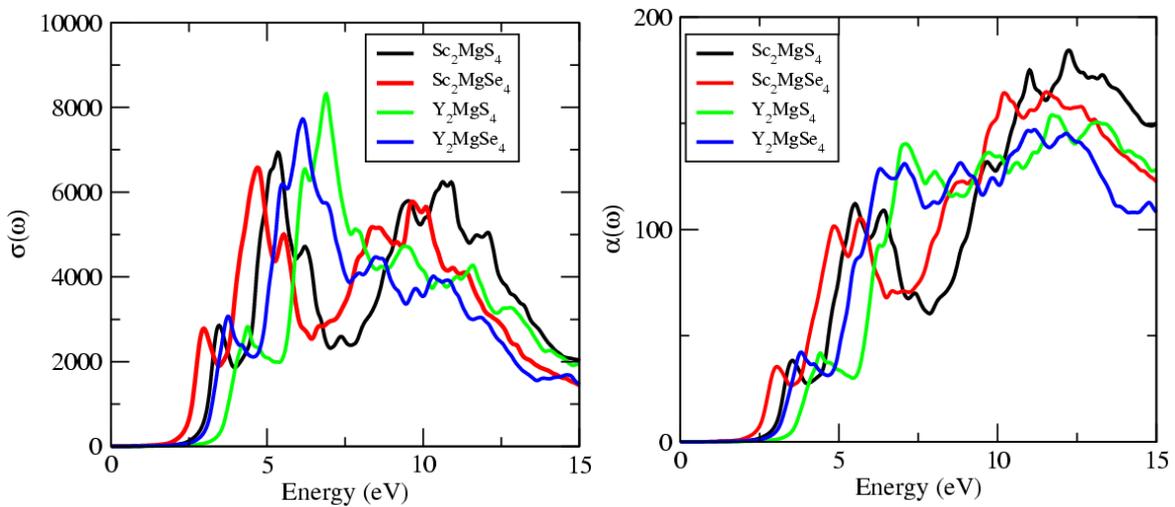

**Fig. 10**. The optical conductivity and absorption coefficient of the compounds $X_2MgZ_4$(X= Sc, Y; Z= S, Se)

**Table 1**: The structural and mechanical parameters of the $X_2MgZ_4$(X= Sc, Y; Z= S, Se) compounds

|  | $V(a.u)^3$ | $a(Å)$ | $B(GPa)$ | $B^p$ | $E_o(Ry.)$ | $C_{11}$ | $C_{12}$ | $C_{44}$ |
|---|---|---|---|---|---|---|---|---|
| $Sc_2MgS_4$ | 2081.84 | 10.73 | 75.55 | 5.0 | -13304.156 | 96.53 | 33.29 | 31.47 |
| $Sc_2MgSe_4$ | 2399.70 | 11.25 | 61.57 | 5.0 | -45800.567 | 83.54 | 24.42 | 23.76 |
| $Y_2MgS_4$ | 2359.47 | 11.18 | 68.57 | 5.0 | -34276.176 | 146.61 | 30.55 | 36.01 |
| $Y_2MgSe_4$ | 2695.27 | 11.69 | 56.95 | 5.0 | -66772.609 | 90.70 | 39.80 | 6.74 |

**Table 2**: The electron and hole effective masses of the $X_2MgZ_4$ (X= Sc, Y; Z= S, Se) compounds

|  | Electron effect mass ($m_e^*$) | Hole effective mass ($m_h^*$) | Energy bandgap (in eV) |
|---|---|---|---|
| $Sc_2MgS_4$ | $0.1948*10^{-31}$ | $0.2580*10^{-31}$ | 2.39 |
| $Sc_2MgSe_4$ | $0.1124*10^{-31}$ | $0.1936*10^{-31}$ | 1.81 |
| $Y_2MgS_4$ | $0.0936*10^{-31}$ | $0.3072*10^{-31}$ | 2.78 |
| $Y_2MgSe_4$ | $0.0828*10^{-31}$ | $0.2448*10^{-31}$ | 2.17 |